# A DWT, DCT AND SVD BASED WATERMARKING TECHNIQUE TO PROTECT THE IMAGE PIRACY

Md. Maklachur Rahman[1]

[1]Department of Computer Science and Engineering, Chittagong University of
Engineering and Technology, Bangladesh
mcr.rahman@gmail.com

## ABSTRACT

*With the rapid development of information technology and multimedia, the use of digital data is increasing day by day. So it becomes very essential to protect multimedia information from piracy and also it is challenging. A great deal of Copyright owners is worried about protecting any kind of illegal repetition of their information. Hence, facing all these kinds of problems development of the techniques is very important. Digital watermarking considered as a solution to prevent the multimedia data.*

*In this paper, an idea of watermarking is proposed and implemented. In proposed watermarking method, the original image is rearranged using zigzag sequence and DWT is applied on rearranged image. Then DCT and SVD are applied on all high bands LH, HL and HH. Watermark is then embedded by modifying the singular values of these bands. Extraction of watermark is performed by the inversion of watermark embedding process. For choosing of these three bands it gives facility of mid-band and pure high band that ensures good imperceptibility and more robustness against different kinds of attacks.*

## 1.INTRODUCTION

In recent years, the increasing amount of applications using digital multimedia technologies has emphasized the need to protect digital multimedia data from pirates. Authentication and information hiding, copyright protection, content identification and proof ownership have also become important issues. To accomplish these issues, watermarking technology is used. Researchers are interested in the field of watermarking because of its significance. These kinds of work in this field have lead to several watermarking techniques such as spatial domain and transform domain. In transform domain it may discrete cosine transform (DCT), discrete wavelet transform (DWT), singular value decomposition (SVD) and their cross relation. Watermarking is a process embedding a piece of information into a multimedia content, such as image, audio and video in such a way that it is imperceptible to a human, but easily detectable by computer. Before the development of digital image watermarking, it was difficult to achieve copyright protection, authentication, data hiding, content identification and proof ownership. But currently it is easy to accomplish these kinds goal using watermarking techniques. So watermarking is very important to us for these kinds of work. Every watermarking algorithm consists of an embedding and extraction process that needs to hide desired information. The embedding and extraction process are described below. Embedded watermark may have several properties such as imperceptibility and robustness. If we cannot distinguish between host image and watermarked image called imperceptibility. Basically imperceptibility depends on similarity between host image and watermarked image. If it difficult to remove or destroy watermark from watermarked image then it said to be robustness. Robustness measures how difficult to remove or destroy watermark from watermarked image. If it is high then robustness is high. DCT based watermarking contains the low frequency information so image contains all information that is similar to the original image. DWT based compression offers scalability so image can be divided into four sub bands in every level of decomposed and by choosing of the sub-band to develop a hybrid watermarking scheme for improving the robustness, imperceptibility and capacity and help to develop a new hybrid method. In the first method (ref. 9) the original image is segmented into blocks then find out spatial frequency of each block select reference image under certain condition then apply DWT, DCT and SVD transformation. In the second method (ref. 3) the original image is segmented into blocks then find out spatial frequency of each block select reference image under certain condition then apply DWT and SVD transformation. In the proposed method, since high band is considered so it fulfills the requirements imperceptibility and robustness. Most of the domain transformation watermarking technique works with DCT, DWT, SVD and their mixing algorithm such as DWT-DCT, DCT-SVD and so on. In this paper we proposed a digital watermarking technique using DWT-DCT and SVD transformation. This method is provided a good imperceptibility and high robustness against various kinds processing attacks. The rest of the paper is organized as follows: Section 2, focuses on overview of transforms for watermarking. Section 3, gives details the proposed methodology and watermarking algorithms. In section 4, gives experimental results and compares. In section 5, conclusion is drawn.

## 2. PRELIMINARIES

As stated earlier that transform domain based watermarking scheme is always a better choice than spatial domain based watermarking scheme. This can be done by using different transformation like DCT, SVD and DWT. In this section, we will briefly describe the DCT, DWT and SVD transformations in below.

### 2.1. Discrete Wavelet Transform (DWT)

The basic idea of DWT in which a one dimensional signal is divided in two parts one is high frequency part and another is low frequency part. Then the low frequency part is split into two parts and the similar process will continue until the desired level. The high frequency part of the signal is contained by the edge components of the signal. In each level of the DWT (Discrete Wavelet Transform) decomposition an image separates into four parts these are approximation image (LL) as well as horizontal (HL), vertical (LH) and diagonal (HH) for detail components. In the DWT decomposition input signal must be multiple of $2^n$. Where, n represents the number of level. To analysis and synthesis of the original signal DWT provides the sufficient information and requires less computation time. Watermarks are embedded in these regions that help to increase the robustness of the watermark. A one level DWT decomposition process is shown in Figure 1.

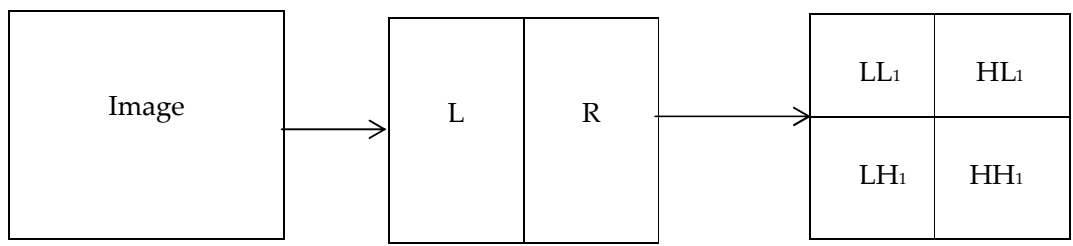

Figure 1. One level DWT decomposition process

### 2.2. Discrete Cosine Transform (DCT)

The DCT is the most popular transform function used in signal processing. It transforms a signal from spatial domain to frequency domain. Due to good performance, it has been used in JPEG standard for image compression. It is a function represents a technique applied to image pixels in spatial domain in order to transform them into a frequency domain in which redundancy can be branded. DCT techniques are more robust compared to spatial domain techniques. Such algorithms are robust against simple image processing operations like adjustment, brightness, blurring, contrast and low pass filtering and so on. But it is difficult to implement and computationally more expensive. The one-dimensional DCT is useful in processing one-dimensional signals such as speech waveforms. For analysis of two-dimensional (2D) signals such as images, we need a 2D version of the DCT. The 2D DCT and 2D IDCT transforms is given by equation 1 and 2.

Formulae of 2-D DCT:

$$F(m,n) = \sum_{i=0}^{N-1}\sum_{j=0}^{N-1} C(m)C(n)f(i,j) \cos\left[\frac{\pi(2i+1)m}{2N}\right] * \cos\left[\frac{\pi(2j+1)m}{2N}\right] \quad (1)$$

Formulae of 2-D inverse DCT:

$$f(i,j) = \sum_{i=0}^{N-1}\sum_{j=0}^{N-1} C(m)C(n)F(m,n) \cos\left[\frac{\pi(2i+1)m}{2N}\right]$$

$$* \cos\left[\frac{\pi(2j+1)m}{2N}\right] \quad (2)$$

Where,

$$C(m), C(n) = \begin{cases} \sqrt{\frac{1}{N}}, & m, n = 0 \\ \sqrt{\frac{2}{N}}, & m, n = 1 \; up \; to \; N-1 \end{cases}$$

## 2.3. Singular Value decomposition (SVD)

The singular value decomposition (SVD) matrix is very useful in computer vision as a decomposition matrix and it is an efficient tool for image transformations. The SVD of a given image $F$ in the form of a matrix is defined as

$$F = USV^T \quad (3)$$

Where, S is the diagonal matrix that is

$$S = \begin{bmatrix} s_1 & 0 & . & 0 & 0 \\ 0 & s_2 & . & 0 & 0 \\ . & . & . & . & . \\ 0 & 0 & . & s_{n-1} & 0 \\ 0 & 0 & . & 0 & s_n \end{bmatrix}$$

And U and V are the orthogonal matrices

$$U^T U = V^T V = I$$

$$VV^T = I$$

$$s_1, s_2, \ldots \ldots s_{n-1}, s_n \geq 0$$

The diagonal elements of matrix **S** are the singular values of matrix **F** and non-negative numbers.

## 3. PROPOSED METHODOLOGY AND WATERMARKING ALGORITHMS

In the proposed watermarking technique, A DWT, DCT and SVD based hybrid watermarking technique is formulated. In this subsection, we have described the watermark embedding and extraction process by using flowchart and algorithmically.

### 3.1. Watermark Embedding Procedure

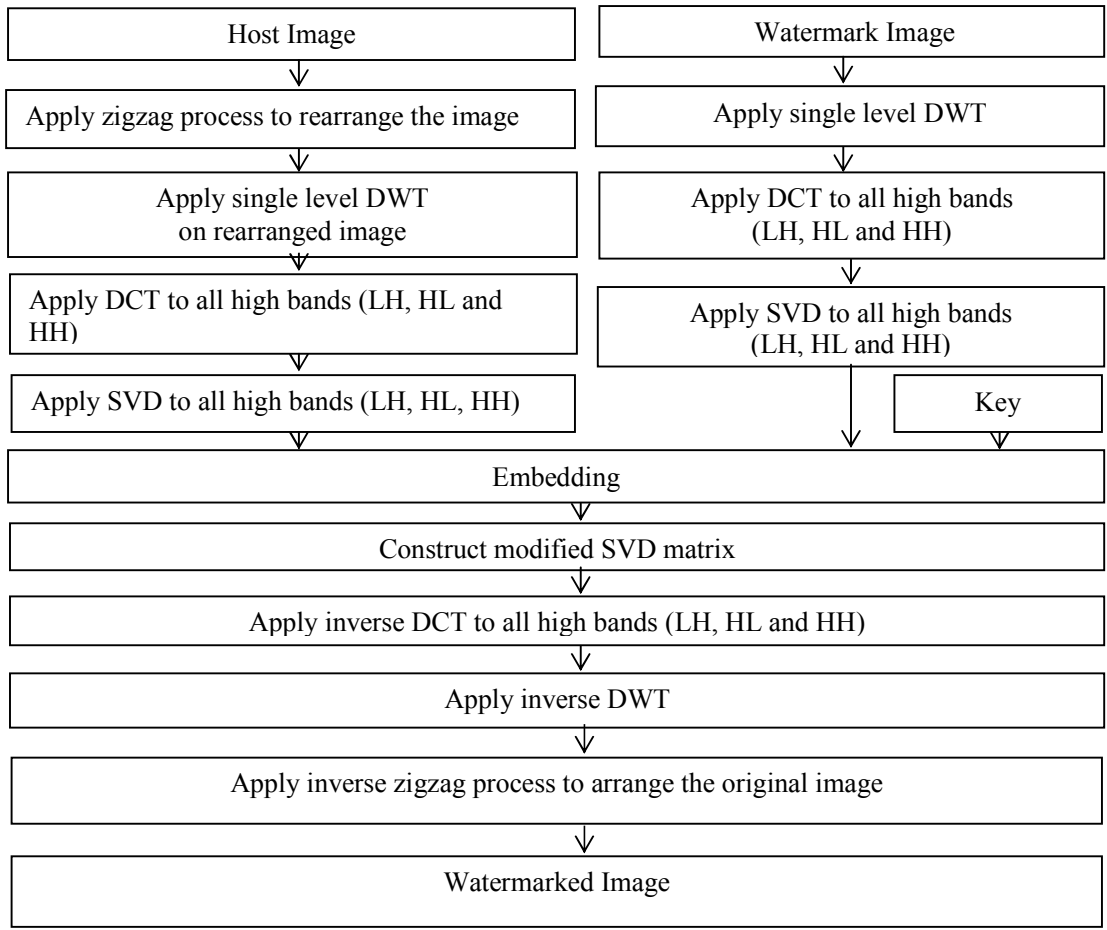

Figure 2. Watermark embedding process

### 3.2. Watermark Extraction Procedure

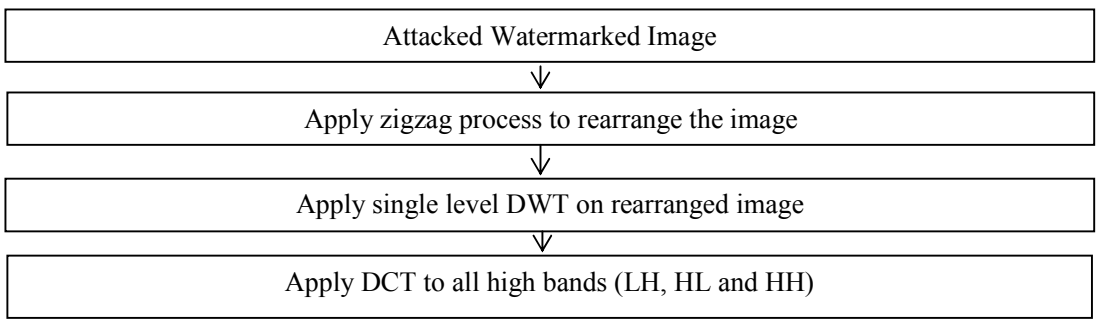

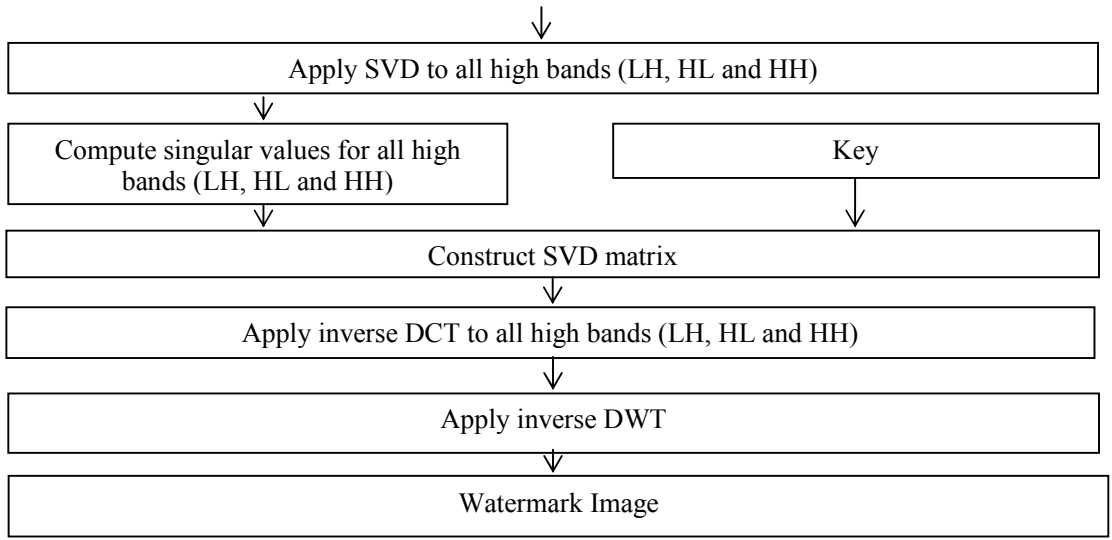

Figure 3. Watermark extraction process

### 3.3. Algorithm: Watermark Embedding

Step 1: Input Host image HI.

Step 2: Rearrange the host image HI by applying zigzag scanning process to get rearranged image RI.

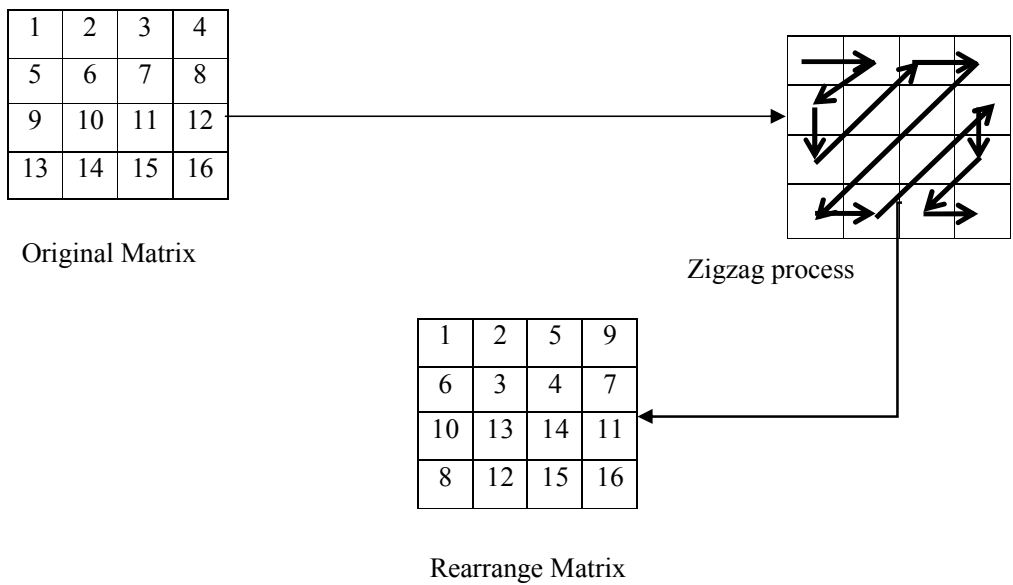

Figure 4. Rearrange a matrix by zigzag process.

Step 3: Apply single level DWT on rearranged image RI to decompose it into four sub-bands LL, HL, LH and HH.

Step 4: Select all high bands LH, HL and HH of RI. Apply DCT to all high bands LH, HL and HH.

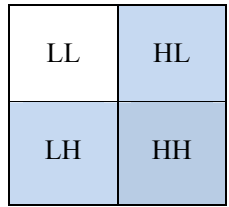

Figure 5. Selected bands for watermarking.

Step 5: Then apply SVD to all high bands LH, HL and HH to get SH1, SH2 and SH3.

Step 6: Input watermark image wi. Apply single level DWT to decompose it into four sub-bands LL1, HL1, LH1 and HH1.

Step 7: Select all high bands LH1, HL1 and HH1 of wi. Apply DCT to all high bands LH1, HL1 and HH1.

Step 8: Then apply SVD to all high bands LH, HL and HH to get SW1, SW2 and SW3.

Step 9: Modify SH1, SH2 and SH3 by using equation $S_{wi} = S_i + \alpha * S_w\ where, i = 1\ upto\ 3$.

Step 10: Construct modified SVD matrix LH11, HL11 and HH11.

Step 11: Apply inverse DCT to all high bands LH11, HL11 and HH11. Apply inverse DWT with LL.

Step 12: Apply inverse zigzag process to arrange the original position of image and finally get watermarked image WI.

### 3.4. Algorithm: Watermark Extraction

Step 1: Input Watermarked image WI.

Step 2: Rearrange the watermarked image WI by applying zigzag scanning process to get rearranged image RI*.

Step 3: Apply single level DWT on rearranged image RI* to decompose it into four sub-bands LL*, HL*, LH* and HH*.

Step 4: Select all high bands LH*, HL* and HH* of RI*. Apply DCT to all high bands LH*, HL* and HH*.

Step 5: Then apply SVD to all high bands LH*, HL* and HH* to get SH1*, SH2* and SH3*.

Step 6: Modify SH1*, SH2* and SH3* by using equation $S_w = (S_{wi} - S_i)/\alpha\ where, i = 1\ upto\ 3$.

Step 7: Construct modified SVD matrix LH1*, HL1* and HH1*.

Step 8: Apply inverse DCT to all high bands LH1*, HL1* and HH1*.

Step 9: Apply inverse DWT to all bands to get watermark image.

## 4. EXPERIMENTAL PERFORMANCE ANALYSIS

The proposed watermarking algorithm is simulated using MATLAB 9 with Processor Intel core 2 duo 2.2 GHz and RAM 2 GB. The proposed watermarking algorithm is tested for the various host and watermark images. Here some results are given. To evaluate the performance of the proposed method, calculate PSNR (Peak Signal to Noise Ratio) and NCC (Normalized Cross Correlation) values. PSNR is widely used to measure imperceptibility between the original image and watermarked image. PSNR is defined by the eqn. (5). The similarity between the original and extract watermark image use to represent how algorithm is robust against noise that is calculated by NCC value. NCC is defined by eqn. (6).

$$PSNR = 10\log_{10}\left(\frac{255^2}{MSE}\right) \tag{5}$$

Where,

$$MSE = \frac{1}{M \times N} \sum_{m=1}^{M} \sum_{n=1}^{N} [I(m,n) - I_w(m,n)]^2$$

$$NCC = \frac{\sum_i \sum_j w(i,j)\, w'(i,j)}{\sum_i \sum_j |w(i,j)|^2} \tag{6}$$

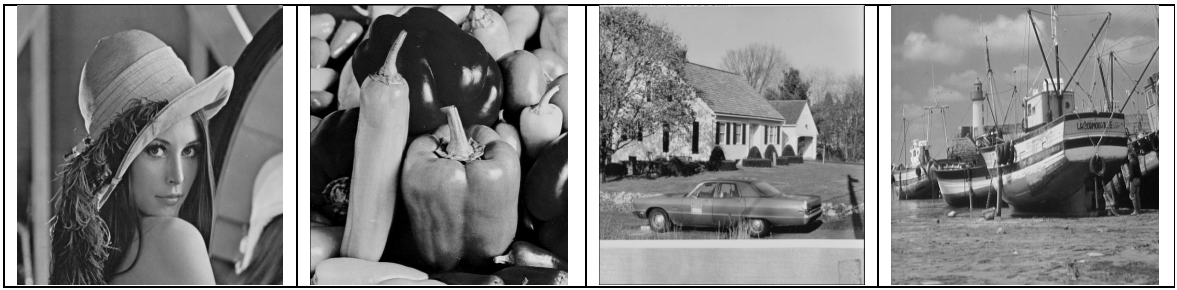

Figure 6. Host images

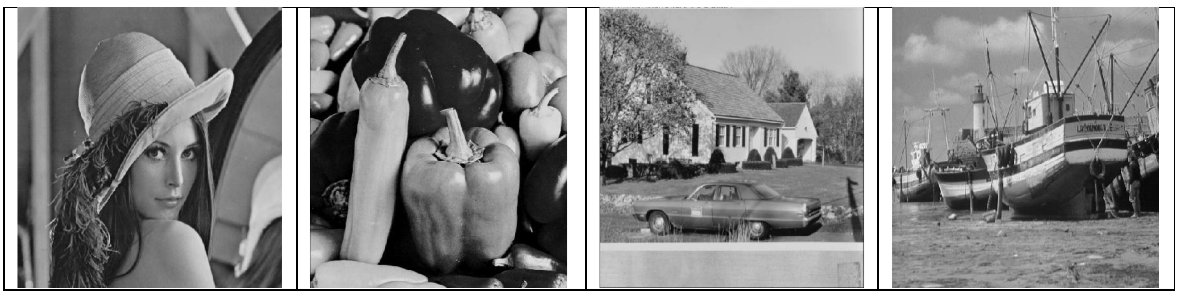

Figure 7. Watermarked images with watermark copyright image

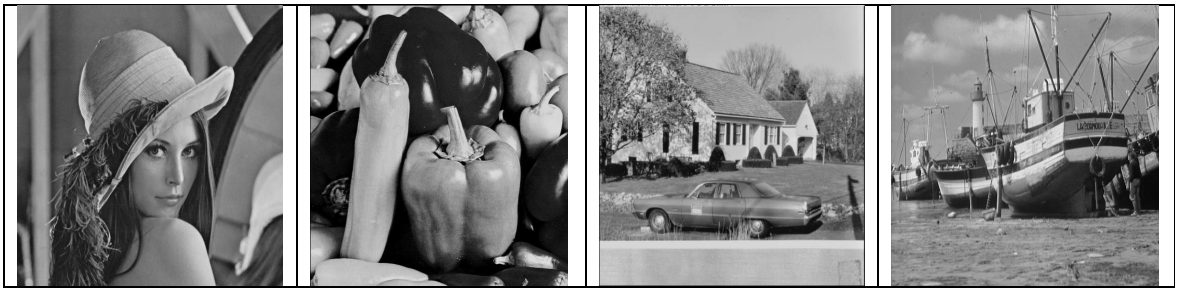

Figure 8. Watermarked images with watermark cameramen image

| PSNR = 40.5935 | PSNR = 40.6685 | PSNR = 40.0854 | PSNR = 34.3189 |
| NCC = 0.9993 | NCC = 0.9991 | NCC = 0.9994 | NCC = 0.9997 |

Figure 9. Extracted copyright watermark Images without applying noise

| PSNR = 36.5698 | PSNR = 37.2255 | PSNR = 36.4139 | PSNR = 32.9252 |
| NCC = 0.9988 | NCC = 0.9979 | NCC = 0.9988 | NCC = 0.9994 |

Figure 10. Extracted CUET logo watermark Images without applying noise

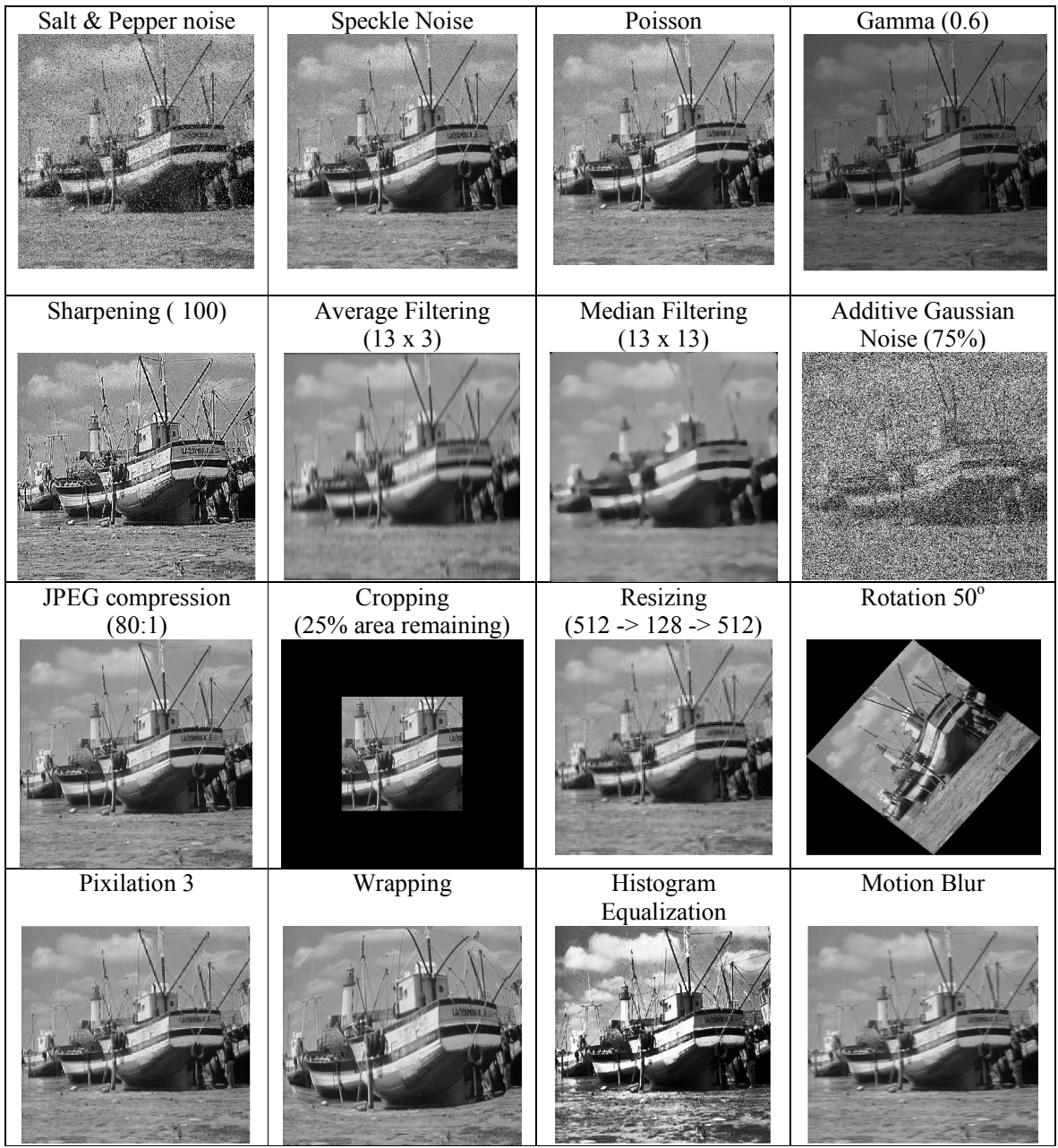

Figure 11. Attacked watermarked image

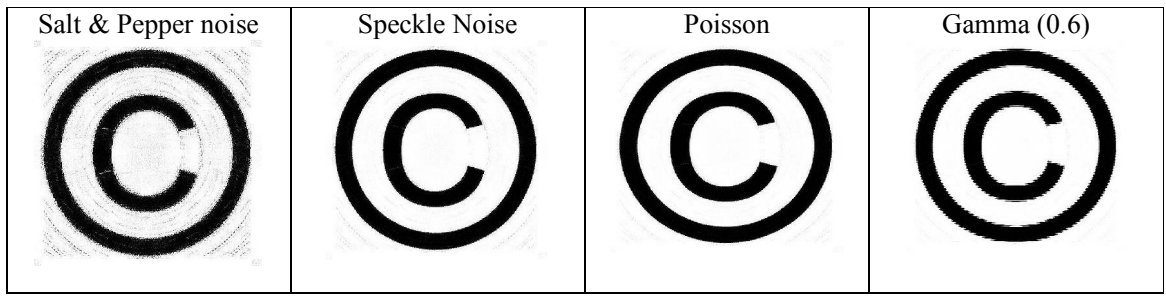

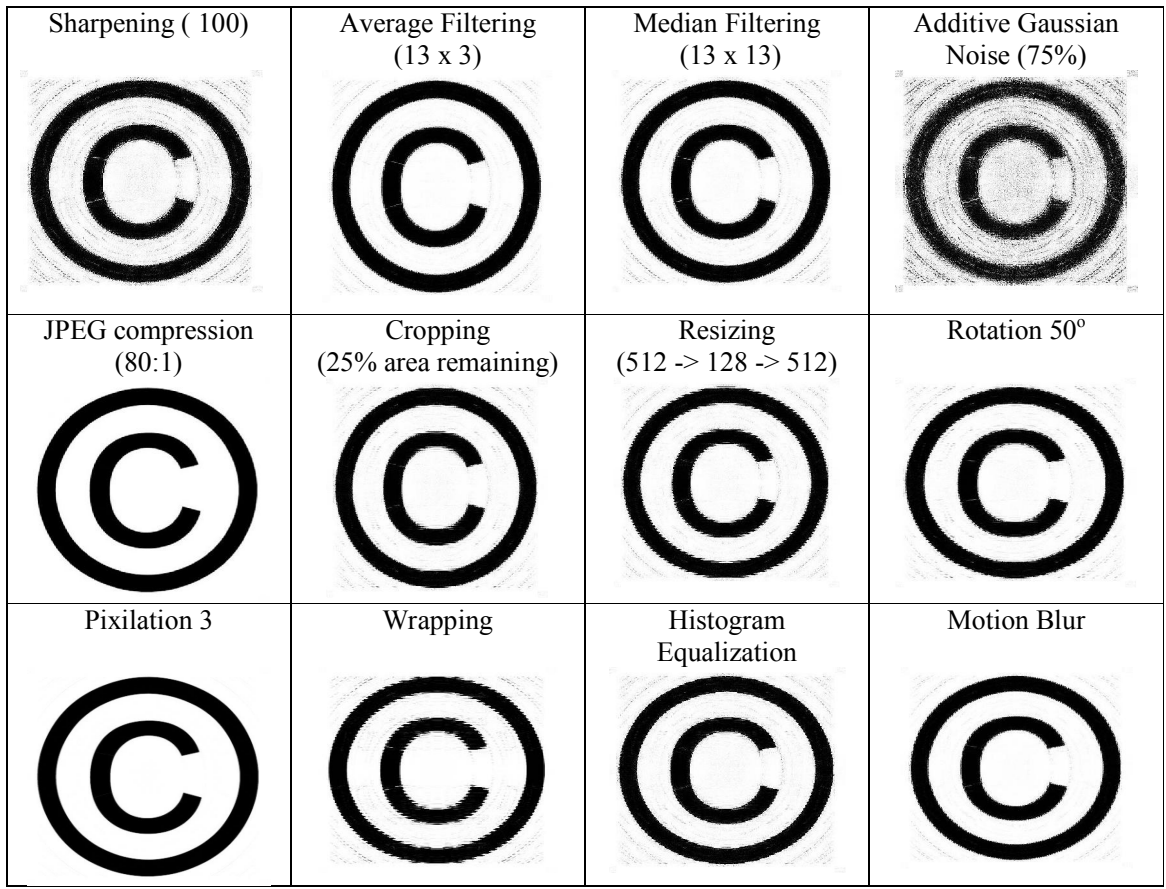

Figure 12. Extracted watermarks from attacked images

Table 1. Performance results in terms of Normalized Cross Correlation (NCC) values

| Attacks | Normalized Cross Correlation (NCC) values | | |
|---|---|---|---|
| | Existing methods | | Proposed Method |
| | DWT-SVD (Ref : 3) | DWT-DCT-SVD (Ref : 9) | DWT-DCT-SVD |
| Average Filtering(13 x 3) | 0.1198 | -0.0928 | **0.9589** |
| Median Filtering (13 x 13) | -0.0852 | -0.0852 | **0.9358** |
| Additive Gaussian Noise (75%) | **0.6749** | **0.6749** | 0.4255 |
| JPEG compression (80:1) | 0.9751 | 0.9751 | **0.9997** |
| Cropping (25% area remaining) | 0.8810 | 0.6120 | **0.9530** |
| Resizing (512 -> 128 -> 512) | 0.2570 | 0.2570 | **0.7391** |
| Rotation 50º | 0.8846 | 0.8846 | **0.9338** |
| Pixilation 3 | 0.0871 | -0.4185 | **0.9983** |
| Wrapping | 0.7299 | -0.4559 | **0.8600** |
| Histogram equalization | **0.9182** | **0.9182** | 0.8416 |
| Motion blur | -0.1854 | -0.0363 | **0.9729** |
| Sharpening | 0.7240 | 0.7500 | **0.7727** |
| Salt & Pepper noise | --------- | --------- | 0.7889 |
| Speckle Noise | --------- | --------- | 0.9873 |
| Poisson | --------- | --------- | 0.9941 |
| Gamma (0.6) | --------- | --------- | **0.9041** |

From Table 1, it is experimental that the proposed DWT-DCT-SVD watermarking algorithm gives more NCC values than the existing DWT-SVD and DWT-DCT-SVD method. That ensures more robustness against different kinds of noise. And Figure 5 and 6 also shows the good PSNR values that ensures more imperceptibility.

## 5. CONCLUSIONS

The proposed watermarking algorithm using DWT, DCT and SVD transformation that contributes more robust in comparison with many watermarking algorithms. The watermarked image quality is good in terms of imperceptibility. In this watermarking algorithm all high bands LH, HL, HH are chosen which cover the mid bands LH, HL and pure high band HH that gives more robust against different kinds of filtering noises and geometric noises. In future, the proposed algorithm can be improved using full band DWT-DCT-SVD and further can be extended to color images and video processing.

**Author**

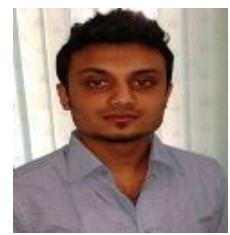

**Md. Maklachur Rahman** will receive B.Sc. degree in Computer Science and Engineering (CSE) from Chittagong University of Engineering and Technology (CUET), Chittagong, Bangladesh in July, 2013. Currently he is a final year student Dept. of CSE. His research interest includes Digital Image Processing, Multimedia Security, Artificial Intelligence, Human Computer Interaction, Digital Watermarking and Software Engineering.